%% file: conf.tex
\documentclass[11pt]{article}
\usepackage{graphicx}
\usepackage{amsmath}

\newcommand{\BABARPubYear}    {02}

\newcommand{\BABARConfNumber} {031}
\newcommand{\SLACPubNumber} {9320}

\input pubboard/babarsym

\newcommand{\kevcc}{\ensuremath{{\mathrm{\,ke\kern -0.1em V\!/}c^2}}\xspace}

\long\def\inst#1{\par\nobreak\kern 4pt\nobreak
    {\it #1}\par\vskip 10pt plus 3pt minus 3pt}

\begin{document}
{\pagestyle{empty}

\begin{flushright}
%BAD 408, Version 10 \\
\babar-CONF-\BABARPubYear/\BABARConfNumber \\
%\babar-PUB-\BABARPubYear/\BABARPubNumber \\
SLAC-PUB-\SLACPubNumber \\
%hep-ex/\LANLNumber \\
July 2002 \\
\end{flushright}

\par\vskip 5cm

% Title of the paper
\begin{center}
\Large \bf 
  Dalitz Plot Analysis of \boldmath{$D^0$} Hadronic Decays
   \boldmath{$D^0 \to K^0 K^- \pi^+$}, \boldmath{$D^0 \to 
\overline{K}^0 K^+ \pi^-$}
 and \boldmath{$D^0 \to \overline{K}^0 K^+ K^-$}
\end{center}
\bigskip

\begin{center}
\large The \babar\ Collaboration\\
\mbox{ }\\
July 27,2002
%\today
\end{center}
\bigskip \bigskip

% Abstract
\begin{center}
\large \bf Abstract
\end{center}
A Dalitz plot analysis of the $D^0$ hadronic decays 
$D^0 \to K^0 K^- \pi^+$, $D^0 \to \overline{K}^0 K^+ \pi^-$ and 
$D^0 \to \overline{K}^0 K^+ K^-$ is presented. 
This analysis is based on a data sample of 22 \invfb collected with the 
\babar\ detector 
at the PEP-II asymmetric-energy B Factory at SLAC
running on or near the \Y4S resonance.
The events are selected from continuum 
$e^+ e^-$ annihilations using the decay $D^{*+} \to D^0 \pi^+$.
Preliminary measurements of the branching fractions of the above hadronic 
decays are obtained.
Preliminary estimates of fractions and phases for resonant and nonresonant
contributions to the Dalitz plot are also presented.
\vfill
\begin{center}
Contributed to the 31$^{st}$ International Conference on High Energy Physics,\\ 
7/24---7/31/2002, Amsterdam, The Netherlands
\end{center}

\vspace{1.0cm}
\begin{center}
{\em Stanford Linear Accelerator Center, Stanford University, 
Stanford, CA 94309} \\ \vspace{0.1cm}\hrule\vspace{0.1cm}
Work supported in part by Department of Energy contract DE-AC03-76SF00515.
\end{center}

\newpage
} % end of pagestyle{empty}

% Input author list file
\input pubboard/authors_ICHEP2002.tex

% The body of the paper starts here
\section{Introduction}
\label{sec:Introduction}

The Dalitz plot analysis is the most complete method of studying
the dynamics of three-body charm decays.  
These decays are
expected to proceed through intermediate resonant two-body modes 
\cite{two} and experimentally
this is the observed pattern with some important exceptions. In the case 
of the decay $D^+ \to K^- \pi^+ \pi^+$~\cite{gobel}, 
for example, the data
can be described with a large ($\approx$ 90\%) nonresonant contribution.
Dalitz plot analyses can provide new information on the resonances that
contribute to observed three-body final states. 
  
In addition, since the intermediate two-body modes are dominated by
light mesons, new information on light meson spectroscopy can be obtained. In particular,
old puzzles related to the parameters and the internal structure of several
light mesons can receive new inputs. 

This paper focuses on the study of three-body 
$D^0$ meson decays\footnote{All references in this 
paper to a specific charged state, unless otherwise
specified, imply also the charge conjugate state.} involving a
$K^0$ (where $K^0 \to \pi^+ \pi^-$) such as

\begin{align}
D^0 &\to \overline{K}^0 \pi^+ \pi^-, \label{eq:1} \\
D^0 &\to K^0 K^- \pi^+, \label{eq:2} \\
D^0 &\to \overline{K}^0 K^+ \pi^-, \label{eq:3} \\
D^0 &\to \overline{K}^0 K^+ K^-. \label{eq:4}
\end{align}
When decays (\ref{eq:2}) and (\ref{eq:3}) are 
combined, they are labelled as $D^0 \to K^0 K \pi$. 
All decays are collectively referred to as $K^0 h^+ h^-$.

\section{The \babar\ Detector and Dataset}
\label{sec:babar}

The data sample used in this analysis consists of 22 \invfb recorded 
with the \babar\ detector at the SLAC \pep2\ storage ring between October 1999
and December 2000. The PEP-II facility operates nominally at the \Y4S 
resonance, 
providing collisions of 9.0 \gev electrons on 3.1 \gev positrons. The data set includes 
19.6 \invfb collected in this configuration
(on-resonance) and 2.4 \invfb collected below the 
$B \Bbar$ threshold (off-resonance).

A more complete overview of the \babar\ detector can be found
elsewhere~\cite{ref:babar}. The following is a brief description of the 
components important for this analysis.
The interaction point is surrounded by a 5-layer 
double-sided silicon vertex tracker (SVT) and a 40-layer drift chamber (DCH) 
filled with a gas mixture of helium and isobutane in a 1.5 T superconducting 
solenoidal magnet. 
In addition to providing precise spatial hits for tracking, the SVT and DCH 
also measure $dE/dx$, which provides
particle identification for low-momentum charged particles.
At higher momenta ($p>0.7$~\gevc) pions and kaons are identified by
Cherenkov radiation observed in the DIRC, a detector designed to 
measure internally reflected Cherenkov light. 
The typical separation between pions and kaons varies from $8 \sigma$ 
at 2 \gevc to 2.5$\sigma$ at 4 \gevc, where $\sigma$ is the 
experimental resolution for the measurement of the Cherenkov angle.

\section{\boldmath Event Selection and $D^0$ Reconstruction}
\label{sec:selection}

\begin{figure}[!htb]
\begin{center}
\includegraphics[height=12cm,width=12cm]{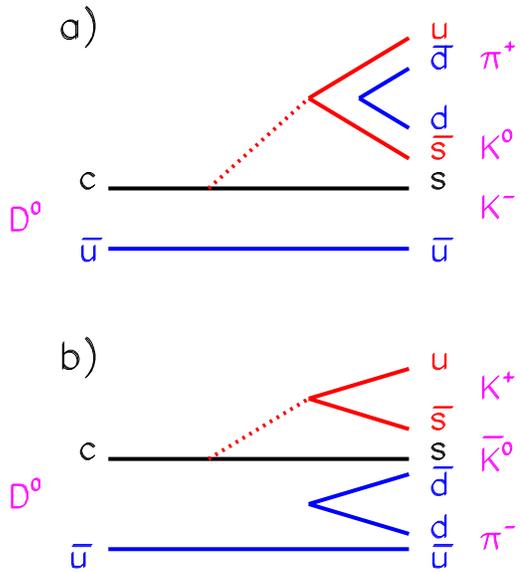}
\caption{Diagrams that contribute to (a) $D^0 \to K^0 K^- \pi^+$  
and (b) $D^0 \to \overline{K}^0 K^+ \pi^-$.}
\label{fig:diag}
\end{center}
\end{figure}

The decay $D^{*+} \to D^0 \pi^+$ is used
to distinguish between $D^0$ and $\overline{D^0}$ and to reduce
background. For example, the Cabibbo-favored decays are
$$D^{*+} \to 
\begin{array}[t]{l} D^0 \pi^+ \\ \to \overline{K}^0 \pi^+ \pi^-, \end{array}$$
$$D^{*-} \to 
\begin{array}[t]{l} \overline{D^0} \pi^- \\ \to K^0 \pi^+ \pi^-. \end{array}$$
The charge of the slow $\pi^{\pm}$ from $D^*$ decay
(referred to as the slow pion) identifies the flavor of the $D^0$ and $K^0$ 
(apart from a small contribution from doubly-Cabibbo-suppressed decays).
The decay channels $K^0 K^- \pi^+$ and $\overline{K}^0 K^+ \pi^-$
 are produced by different decay diagrams, as shown
in Fig.~\ref{fig:diag}; therefore their rates are not expected to be the same.

$D^0 \to K^0 h^+ h^-$ candidates are reconstructed from $K^0_S \to \pi^+ \pi^-$
candidates plus two charged tracks, each with at least 12 hits in the
DCH, and a track with momentum smaller than 0.6 \gevc with
at least 6 hits in the SVT. In addition, tracks are 
required to have transverse momentum $p_T>$ 100 MeV/$c$ and, 
except for the $K^0$ decay pions,
to point back to the nominal interaction
point within 1.5 cm transverse to the beam and 3 cm along the beam direction.

$K^0$ particles are reconstructed by kinematically fitting all pairs of 
positive and negative tracks.
Reconstructed $K^0$ candidates are then fit to a common vertex with all 
remaining combinations of
pairs of positive and negative tracks. 
Fake $K^0$ candidates are removed requiring a flight distance of 0.4 cm 
with respect to the candidate $D^0$ vertex. 
The $D^0$ candidate is then combined with all the slow pion candidates, 
which are refitted to a common vertex constrained to be located 
in the interaction region. In all cases a fit probability cut at 0.1 \% 
is applied.  

In order to reduce the combinatorial background, $D^0$ particles originating 
from $B$ decays are rejected by requiring that the
center of mass momentum of the $D^0$ candidate is greater than 2.2 \gevc. 

Two different particle identification
algorithms (PID) are used for kaon identification, 
both of which take advantage of the measurement of $dE/dx$ in the
tracking detectors and a measurement of the Cherenkov angle in the
DIRC. 
The first employs less stringent requirements and has an efficiency
above 95\% for kaons with momentum less than 3~\gevc
and $\pi$ misidentification probability of about 20\%.
The second algorithm is more strict and has kaon
identification efficiencies of 70\% to 90\% in the same
momentum range but with $\pi$ misidentification probabilities
below 7\%. The more permissive $K$ identification algorithm
is used for the measurement of branching ratios, where
more uniform acceptance is desirable, and the tighter $K$ algorithm
is used for the Dalitz plot analyses, where background reduction is more
important.

\begin{figure}[!htb]
\begin{center}
\includegraphics[height=18cm,width=10cm]{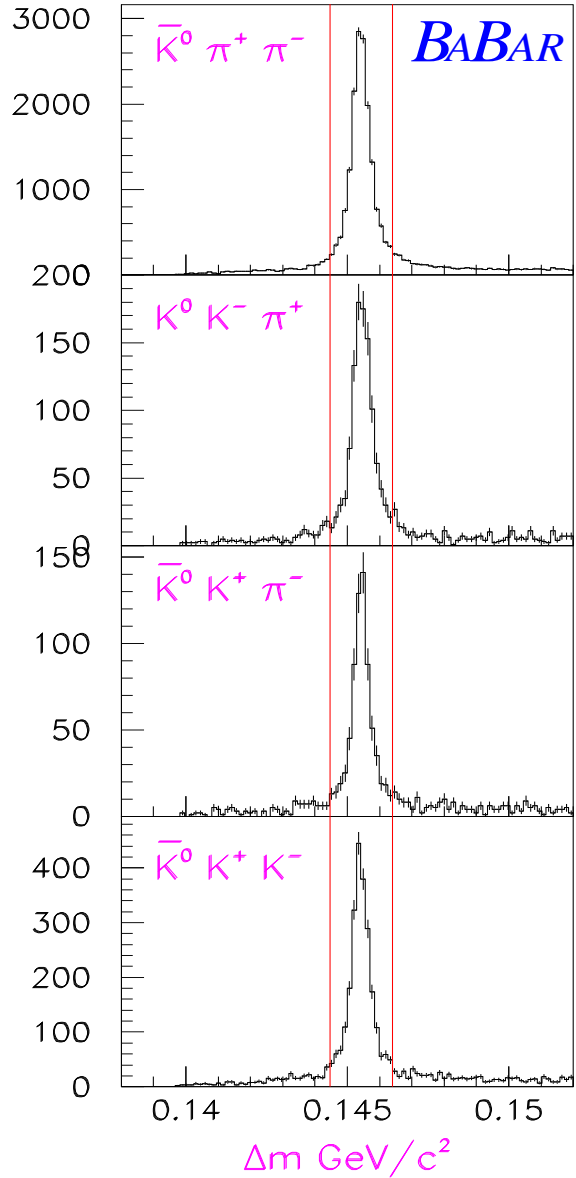}
\caption{$\Delta m$ distributions for $D^0 \to K^0 h^+ h^-$ candidates,
for events in which the
$K^0 h^+ h^-$  invariant mass is within 2.5 $\sigma$ of the $D^0$ mass.
The lines represent 
the range of $\Delta m$ used to select the $D^0$ candidates.}
\label{fig:deltam}
\end{center}
\end{figure}

\begin{figure}[!htb]
\begin{center}
\includegraphics[height=16cm,width=16cm]{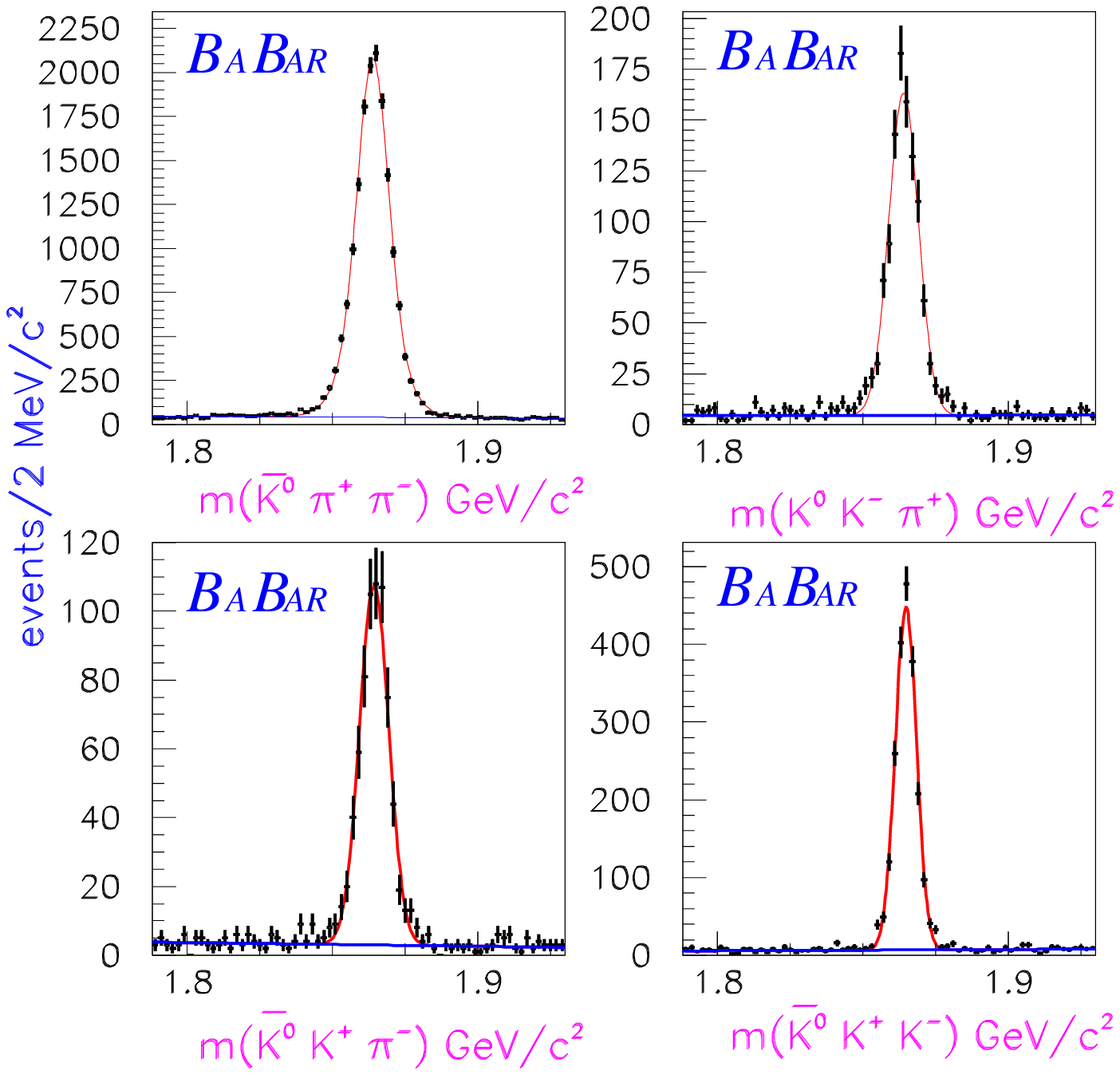}
\caption{$K^0 h^+ h^-$ mass distributions for events in which
$\Delta m$ is within 969 \kevcc of the mean
$\Delta m$ value for signal events.}
\label{fig:d0}
\end{center}
\end{figure}

Each $D^0$ sample is characterized by the distributions of two
variables, the invariant mass of the candidate $D^0$, and the difference
between the invariant masses of the $D^*$ and $D^0$ candidates,
$$\Delta m = m(K^0 h^+ h^- \pi_S^+) - m(K^0 h^+ h^-),$$
where $\pi^+_S$ is the slow pion. 
The distribution of $\Delta m$
for those candidates that fall within 2.5 $\sigma$ of the
$D^0$ mass peak is shown in Fig.~\ref{fig:deltam}. 
A strong $D^*$ signal is apparent.
Fits to these
distributions produce consistent means and widths for the four
channels ($\sigma=(323 \pm 10)$ \kevcc in the case
of decay channel $\overline{K}^0\pi^+\pi^-$).
The $K^0 h^+ h^-$ mass distribution for candidates that fall within 
$\pm969$~\kevcc (corresponding to 3 $\sigma$) from the central value
 of the $\Delta m$ distribution is shown in Fig.~\ref{fig:d0}. 
 

\section{Efficiency}
\label{sec:Efficiency}

The efficiency for the decays in the four $D^0$ samples is
determined from a sample of Monte Carlo events in which each 
decay mode is generated
according to uniform phase space (such that the Dalitz plot is uniformly
populated). These events are passed through a full detector
simulation based on GEANT3 \cite{geant} and subjected to the same
reconstruction and event selection as the data. The distribution
of these events in the Dalitz plot after selection is used to
determine the reconstruction efficiency. Typical Monte Carlo samples used to 
compute these efficiencies consist of 400 $\times 10^3$ events.

\begin{figure}[!htb]
\begin{center}
\includegraphics[height=12cm,width=12cm]{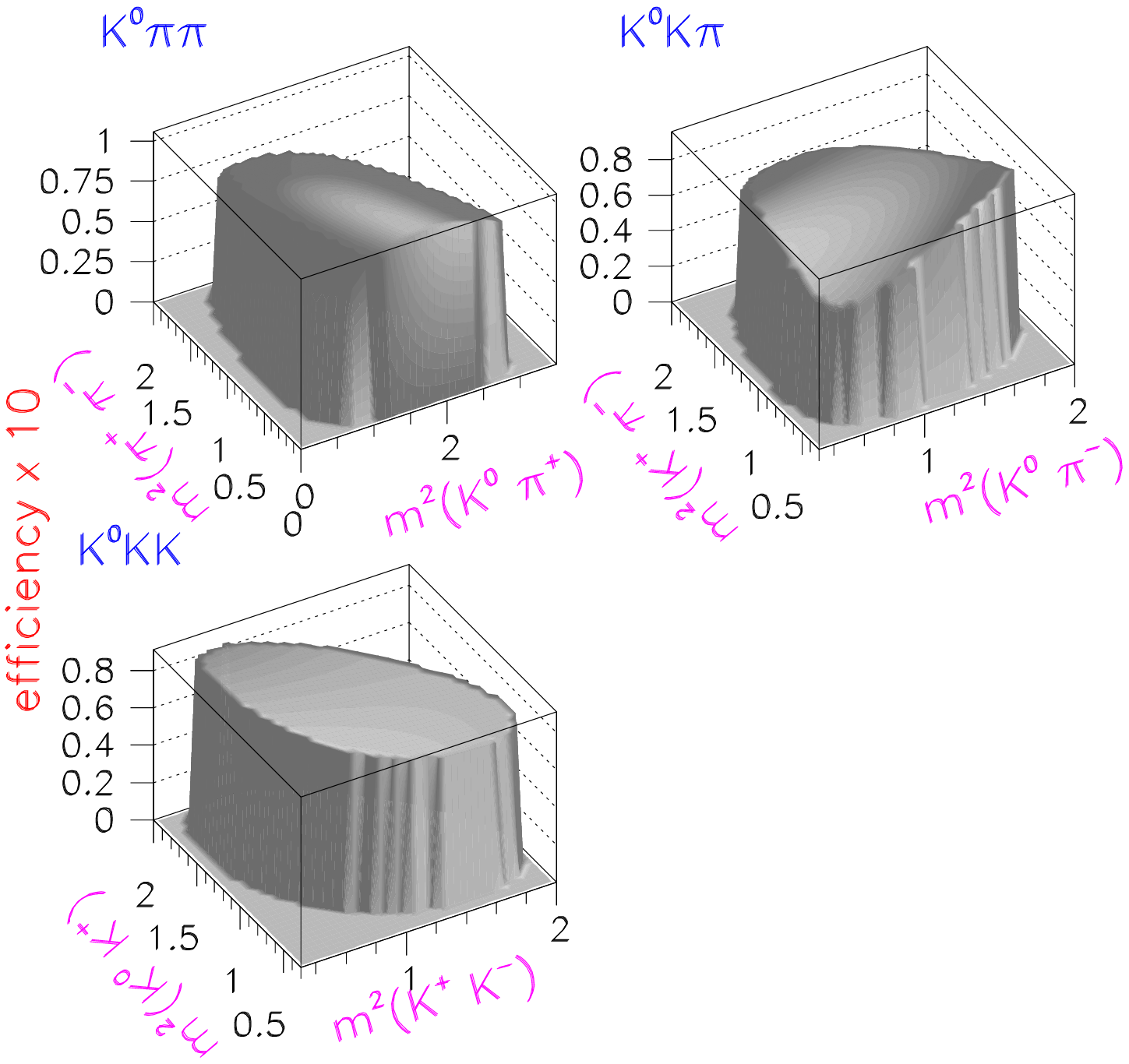}
\caption{Dalitz plot efficiency ($\times$ 10) for $D^0 \to \overline{K}^0 \pi^+ \pi^-$,
$D^0 \to K^0 K \pi$ and $D^0 \to \overline{K}^0 K^+ K^-$. 
} 
\label{fig:acc}
\end{center}
\end{figure}

This Dalitz distribution is divided into small cells and fit to
a third-order polynomial in two dimensions.
Cells with less than 50 events are ignored in the fit.
The resulting $\chi^2$ per number of degrees of freedom ($\chi^2/NDF$)
is typically 1.1. The fitted efficiencies are shown in Fig.~\ref{fig:acc}.
The average over the Dalitz plot ranges from 7.3\% to 8.7\%, 
depending on the decay mode.

\section{Branching Fractions}

Since all four decay channels $K^0 h^+ h^-$ have similiar topologies,
the ratios of branching fractions, calculated relative to the 
$\overline{K}^0\pi^+ \pi^-$ decay mode, are expected to have relatively
small systematic uncertainties.
These ratios are evaluated as
$$R = \frac{\sum_{x,y} \frac{N_i(x,y)}{\epsilon_i(x,y)}}{\sum_{x,y}\frac{N_1(x,y)}{\epsilon_1(x,y)}},$$
where $N_i(x,y)$ represents the number of events measured for channel $i$
and $\epsilon_i(x,y)$ is the corresponding efficiency in a given Dalitz 
plot cell $(x,y)$ .

In order to obtain the yields and measure the relative branching fractions,
the $K^0 h^+ h^-$ mass distributions are fit with a Gaussian function
and a linear background. 
The number of signal events is calculated as the difference between the
total number of events within $3 \sigma$ of the $D^0$ mass from the fit and 
the integrated linear background function in the same mass range.
The results obtained with this 
procedure are summarized in Table~\ref{tab:d0}.
\begin{table}[htbp]
\caption{
Results from the fits to the $K^0 h^+ h^-$ mass distributions.
The fits are performed using a 
linear background and a Gaussian function for the signal.}
\label{tab:d0}
\begin{center}
\begin{tabular}{lccr}
\hline
Channel & mass (MeV/$c^2$) & $\sigma$ (MeV/$c^2$) & signal events\cr
\hline
& & & \\
$D^0 \to \overline{K}^0 \pi^+ \pi^- $ & 1863.9 $\pm$ 0.5 & 6.29 $\pm$ 0.05 & 15 279 $\pm$ 129\cr
$D^0 \to K^0 K^- \pi^+$  & 1863.9 $\pm$ 0.2 & 5.1 $\pm$ 0.2 & 1 165 $\pm$ 40 \cr
$D^0 \to \overline{K}^0 K^+ \pi^-$  & 1864.3 $\pm$ 0.2 & 5.0 $\pm$ 0.2 & 805 $\pm$ 35 \cr
$D^0 \to \overline{K}^0 K^+ K^- $ & 1864.7 $\pm$ 0.8 & 3.6 $\pm$ 0.1 & 2 109 $\pm$ 48\cr
\hline
\end{tabular}
\end{center}
\end{table}

Systematic errors take into account the different algorithms used for 
particle identification and the way in which the Dalitz plot efficiency 
is used to correct the data.
For example, the full data set is divided into two
samples corresponding to different 
DCH high voltage operating points and the branching ratios are computed 
separately for the two samples.  Differences in the branching ratios are
used as an estimate of systematic uncertainties.
The contributions to the systematic errors 
are summarized in Table~\ref{tab:sys}. 
The total error is obtained by adding in quadrature the single contributions.
\begin{table}[htbp]
\caption{
Contributions to the systematic error on the ratio of branching fractions relative to 
$D^0 \to \overline{K}^0 \pi^+ \pi^-$ (\%) from different
sources. 
}
\label{tab:sys}
\begin{center}
\vskip -0.2cm
\begin{tabular}{lcccc}
\hline
Channel & PID & Efficiency correction & DCH voltage & Total \cr
\hline
$D^0 \to K^0 K^- \pi^+ $ & 0.31 &  0.14  & 0.45  & 0.56 \cr
$D^0 \to \overline{K}^0 K^+ \pi^- $ & 0.18 & 0.04 & 0.37 & 0.41 \cr
$D^0 \to \overline{K}^0 K^+ K^- $ & 0.19 & 0.09 & 0.17 & 0.27 \cr
\hline
\end{tabular}
\end{center}
\end{table}

The method for measuring the branching ratios is 
checked with a different
fully inclusive
$e^+ e^- \to \bar c c$ Monte Carlo sample in which the $D^0$ mesons
decay according to the PDG~\cite{pdg} branching fractions. 
The Monte Carlo events are subjected
to the same reconstruction, event selection and analysis as the real
data. The size of this Monte Carlo sample is comparable 
to that of the data sample. 
The results are found to be statistically consistent
with the branching ratios
assumed in the Monte Carlo generation.

The measured preliminary ratios of branching fractions are shown
in Table~\ref{tab:exp} and 
are compared with those measured by other experiments.

\begin{table}[htbp]
\caption{Ratio of branching fractions relative to 
$D^0 \to \overline{K}^0 \pi^+ \pi^-$ (\%). Note that ARGUS
did not separate $K^0 K^- \pi^+$ from $\overline{K}^0 K^+ \pi^- $.
}
\label{tab:exp}
\begin{center}
\vskip -0.2cm
\begin{tabular}{lcccc}
\hline
Channel & \babar\ & CLEO~\cite{cleo1} & ARGUS~\cite{argus1} & E691~\cite{e691} \cr
\hline
$D^0 \to K^0 K^- \pi^+ $ & 8.32 $\pm$ 0.29 (stat) $\pm$ 0.56(syst) & 10.8 $\pm$ 1.9 & 
16.0 $\pm$ 3.0 & 12.4 $\pm$ 4.7 \\
$D^0 \to \overline{K}^0 K^+ \pi^- $ & 5.68 $\pm$ 0.25(stat) $\pm$ 0.41(syst) & 9.8 $\pm$ 2.0 & & 7.8 $\pm$ 3.9 \\
$D^0 \to \overline{K}^0 K^+ K^- $ & 16.30 $\pm$ 0.37(stat) $\pm$ 0.27(syst) & 17.0 $\pm$ 2.2 & & \\
\hline
\end{tabular}
\end{center}
\end{table}

\section{Dalitz Plot Analysis Method}

An unbinned maximum likelihood fit is performed
for the decay modes
$D^0 \to K^0 K^- \pi^+$, $D^0 \to \overline{K}^0 K^+ \pi^-$ and 
$D^0 \to \overline{K}^0 K^+ K^-$,
in order to use the distribution of events 
in the Dalitz plot to determine the relative amplitudes and phases 
of intermediate resonant and nonresonant states.
 
Following the method used by ARGUS~\cite{argus} and CLEO~\cite{cleo}, 
the likelihood function is written
in the following way:

\begin{equation}
L = \beta \cdot G(m) \frac {\sum_{i,j} c_i c_j^* A_i A_j^*}
{\int{\sum_{i,j} c_i c_j^*A_i A_j^* \epsilon(m_x^2,m_y^2)} dm_x^2 dm_y^2} + \frac{(1 - \beta)}{\int \epsilon(m_x^2,m_y^2) dm_x^2 dm_y^2}. 
\label{eq:5}
\end{equation}
In this expression, $\beta$ represents the fraction of signal obtained from the fit to
the mass spectrum and $\epsilon(m_x^2,m_y^2)$ is the Dalitz plot
efficiency. $G(m)$ is a Gaussian function describing the $D^0$ lineshape 
normalized within the
$2 \sigma$ cut used to perform the Dalitz plot analysis. 
It is assumed that the background events, described by the second term
in Eq.~\ref{eq:5}, uniformly populate the Dalitz plot.
This assumption is 
verified by examining events in the $D^0$ side bands.
The output from the fit is the set of
complex coefficients $c_i$.

In Eq.~5, the integrals are computed using Monte Carlo events taking into
account the Dalitz plot efficiency.
The branching fraction for the resonant or nonresonant contribution $i$
is defined by the following expression:
$$f_i = \frac {|c_i|^2 \int |A_i|^2 dm_x^2 dm_y^2}
{\sum_{j,k} c_j c_k^* \int A_j A_k^* dm_x^2 dm_y^2}. $$
The fractions $f_i$ do not necessarily add up to 1 because of interference
between amplitudes. The errors in the 
fractions are evaluated by propagating the 
full covariance matrix obtained
by the fit.

The phase of each amplitude is measured with respect to the mode with
the largest amplitude.
The amplitudes $A_i$ are each represented by the product of a complex 
Breit-Wigner function and an angular function 
(the Zemach tensors~\cite{zemach}):
$$ A = BW(m) \times T (\Omega). $$
The Breit-Wigner function includes Blatt-Weisskopf form factors~\cite{dump}.
The $f_0(980)$ and $a_0(980)$ resonances is described using  
coupled-channel Breit-Wigner functions with parameters taken
from the CERN/WA76~\cite{me} and the 
Crystal Barrel~\cite{a0} experiments, respectively.
The parameters of the $\phi(1020)$ meson are extracted from the
data ($\Gamma = (4.3 \pm 0.3)$ MeV/$c^2$) since the apparent width is affected by
the experimental resolution.
The resonance parameters of the $K^*_0(1430)$ are $m=1.435$ \gevcc and 
$\Gamma$=279  \gevcc from a recent reanalysis of elastic $K \pi$ scattering
data from the
LASS experiment~\cite{david}. 
The nonresonant contribution (N.R.) is represented by a constant
term with a free phase.

The fit quality is evaluated in the following way. The Dalitz plot
is re-binned into $N_{cells}$ cells
grouping together bins with small event yields. 
Then a Monte Carlo sample is generated with parameters from the fit to data. 
This distribution is used
to evaluate a $\chi^2$ for the Dalitz plot. 
The $\chi^2/N_{cells}$ is displayed in Table~\ref{tab:all} 
for the fit results shown in Tables~\ref{tab:res_k0skpia},
\ref{tab:res_k0skpib},
and \ref{tab:res_k0skk},
together with
the event yield and purity for each of 
the three channels used in the Dalitz plot
analyses.

\begin{table}[htbp]
\caption{
Event yields, purity and $\chi^2/N_{cells}$ for the three channels in the
Dalitz plot analyses.
}
\label{tab:all}
\begin{center}
\begin{tabular}{|l|c|c|c|}
\hline
\bf $D^0$ Decay mode & Events & Purity (\%) & $\chi^2/N_{cells}$ \\
\hline
$D^0 \to K^0 K^- \pi^+$ & 1008 & 95.5 $\pm$ 0.4 & 46/44 \\
$D^0 \to \overline{K}^0 K^+ \pi^-$ & 659 & 95.5 $\pm$ 0.4 & 25/29 \\
$D^0 \to \overline{K}^0 K^+ K^-$ & 1957 & 97.5 $\pm$ 0.2 & 98/59 \\
\hline
\end{tabular}
\end{center}
\end{table}

Systematic errors on the fitted fractions are evaluated by making 
different assumptions in the fits --- for example,
we have performed fits with the Blatt-Weisskopf terms set to 1, 
and  with uniform efficiency across the Dalitz plot.

\section{\boldmath $D^0 \to K^0 K^- \pi^+$ Dalitz Plot Analysis}

\begin{figure}[!htb]
\begin{center}
\includegraphics[height=12cm,width=12cm]{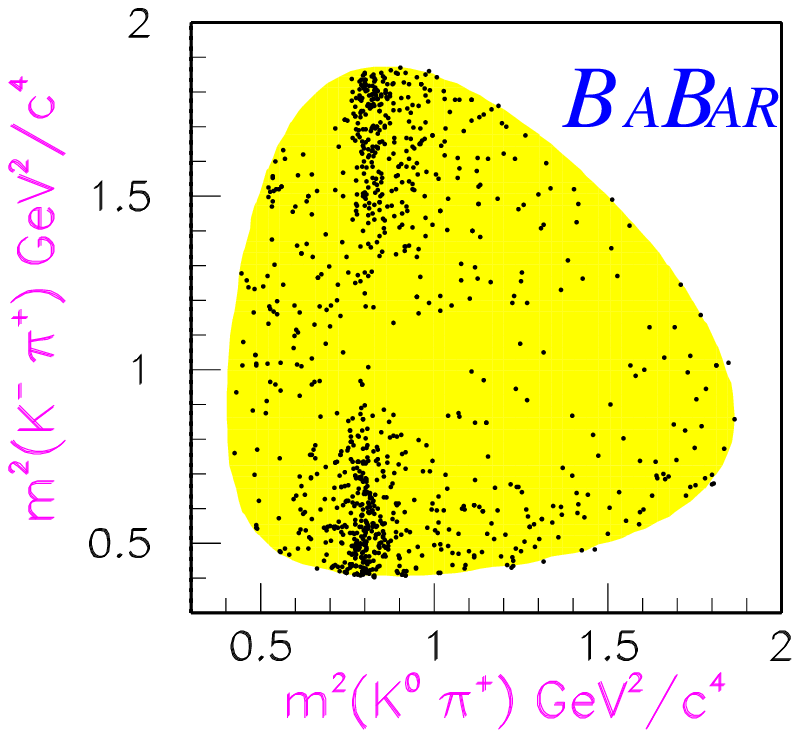}
\caption{Dalitz plot of $D^0 \to K^0 K^- \pi^+$.
} 
\label{fig:dalitz_k0skpi_1}
\end{center}
\end{figure}

\begin{figure}[!htb]
\begin{center}
\includegraphics[height=8cm,width=16cm]{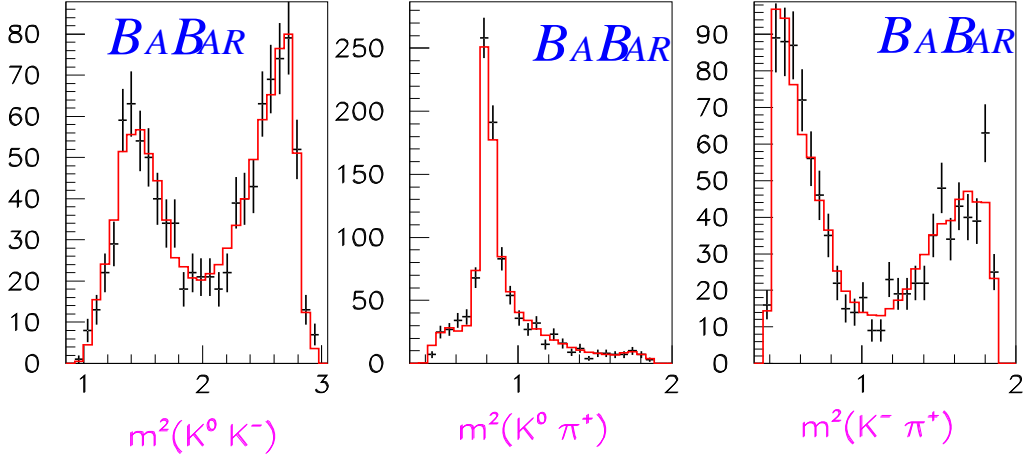}
\caption{Dalitz plot projections for $D^0 \to K^0 K^- \pi^+$. The data are
represented with error bars; the histogram is 
a projection of the fit described in the text.}
\label{fig:pro_k0skpi1}
\end{center}
\end{figure}

The Dalitz plot for $D^0 \to K^0 K^- \pi^+$ candidates
 is shown in Fig.~\ref{fig:dalitz_k0skpi_1} and
its projections are shown in Fig.~\ref{fig:pro_k0skpi1}. The presence of a 
strong $K^{*+}(892)$ resonance can be observed.
Table~\ref{tab:res_k0skpia} shows the list of intermediate final states and the 
fitted fractions for this decay channel.  
\begin{table}[htbp]
\caption{
Preliminary results from the Dalitz plot analysis of $D^0 \to K^0 K^- \pi^+$.
}
\label{tab:res_k0skpia}
\begin{center}
\vskip -0.2cm
\begin{tabular}{lcc}
\hline
 Final state & Fraction (\%) & Phase (degrees)\cr
\hline
& & \\
 $\Kbar^{*0}_0(1430) K^0$ &4.8 $\pm$ 1.4 $\pm$ 1.6& 52 $\pm$ 27\\
 $\Kbar^{*0}_1(892) K^0$ &0.8 $\pm$ 0.5 $\pm$ 0.1 & 175 $\pm$ 22\\
 $\Kbar^{*0}_1(1680) K^0$ & 6.9 $\pm$ 1.2 $\pm$ 1.0& --169 $\pm$ 16 \\
 $\Kbar^{*0}_2(1430) K^0$ &2.0 $\pm$ 0.6 $\pm$ 0.1& 51 $\pm$ 18\\
 $K^{*+}_0(1430) K^-$ & 13.3 $\pm$ 3.5 $\pm$ 3.9&--41 $\pm$ 25\\
 $K^{*+}_1(892) K^-$ & 63.6 $\pm$ 5.1  $\pm$ 2.6 & 0\\
 $K^{*+}_1(1680) K^-$ &  15.6 $\pm$ 3.0 $\pm$ 1.4 &--178 $\pm$ 10 \\
 $K^{*+}_2(1430) K^-$ & 13.8 $\pm$ 2.6 $\pm$ 7.9&--52 $\pm$ 7 \\
 $a_0^-(980) \pi^+$ & 2.9 $\pm$ 2.3 $\pm$ 0.7&--100 $\pm$ 13\\
 $a_0^-(1450) \pi^+$ & 3.1 $\pm$ 1.9 $\pm$ 0.9 & 31 $\pm$ 16\\
 $a_2^-(1310) \pi^+$ & 0.7 $\pm$ 0.4 $\pm$ 0.1 &--149 $\pm$ 27\\
 N.R. & 2.3 $\pm$ 0.5 $\pm$ 5.6  &--136 $\pm$ 23 \\
& & \\
\hline
Sum &  130 $\pm$ 8 & \\
\hline
\end{tabular}
\end{center}
\end{table}

The results from the fit, shown in Table~\ref{tab:res_k0skpia}, confirm that 
the channel is dominated by $K^*$ resonances in the charged mode. Contributions
from neutral $K^*$ as well as contributions from $a_0$ resonances are small
or consistent with zero. The nonresonant contribution appears to be negligible.

Recently, in order to have a better description of the Dalitz plot of 
$D^+ \to K^- \pi^+ \pi^+$, experiment
E791~\cite{gobel} introduced in the analysis 
the old postulated $K\pi$ S-wave resonance $\kappa$~\cite{jaffe}.
The best parameters found were $M_{\kappa}= (797 \pm 19 \pm 42$) \mevcc
and $\Gamma_{\kappa}= (410 \pm 43 \pm 85$) \mevcc.
Including a contribution from $\kappa^+ K^-$ in the present analysis
results in an increase of
$2 \log L$ of 7.8 for 2 more parameters. The resulting fraction for
this contribution is (15 $\pm$ 12)\%. The large error does not allow us to confirm
the existence of this final state.

\section{\boldmath $D^0 \to \overline{K}^0 K^+ \pi^-$ Dalitz Plot Analysis}

\begin{figure}[!htb]
\begin{center}
\includegraphics[height=12cm,width=12cm]{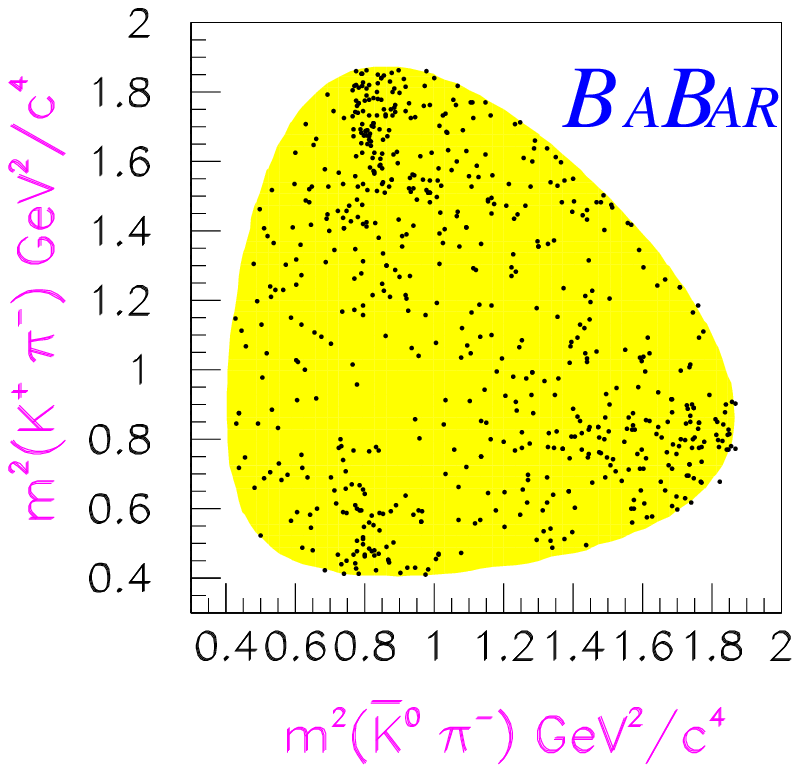}
\caption{Dalitz plot of $D^0 \to \overline{K}^0 K^+ \pi^-$.
} 
\label{fig:dalitz_k0skpi_2}
\end{center}
\end{figure}

\begin{figure}[!htb]
\begin{center}
\includegraphics[height=8cm,width=16cm]{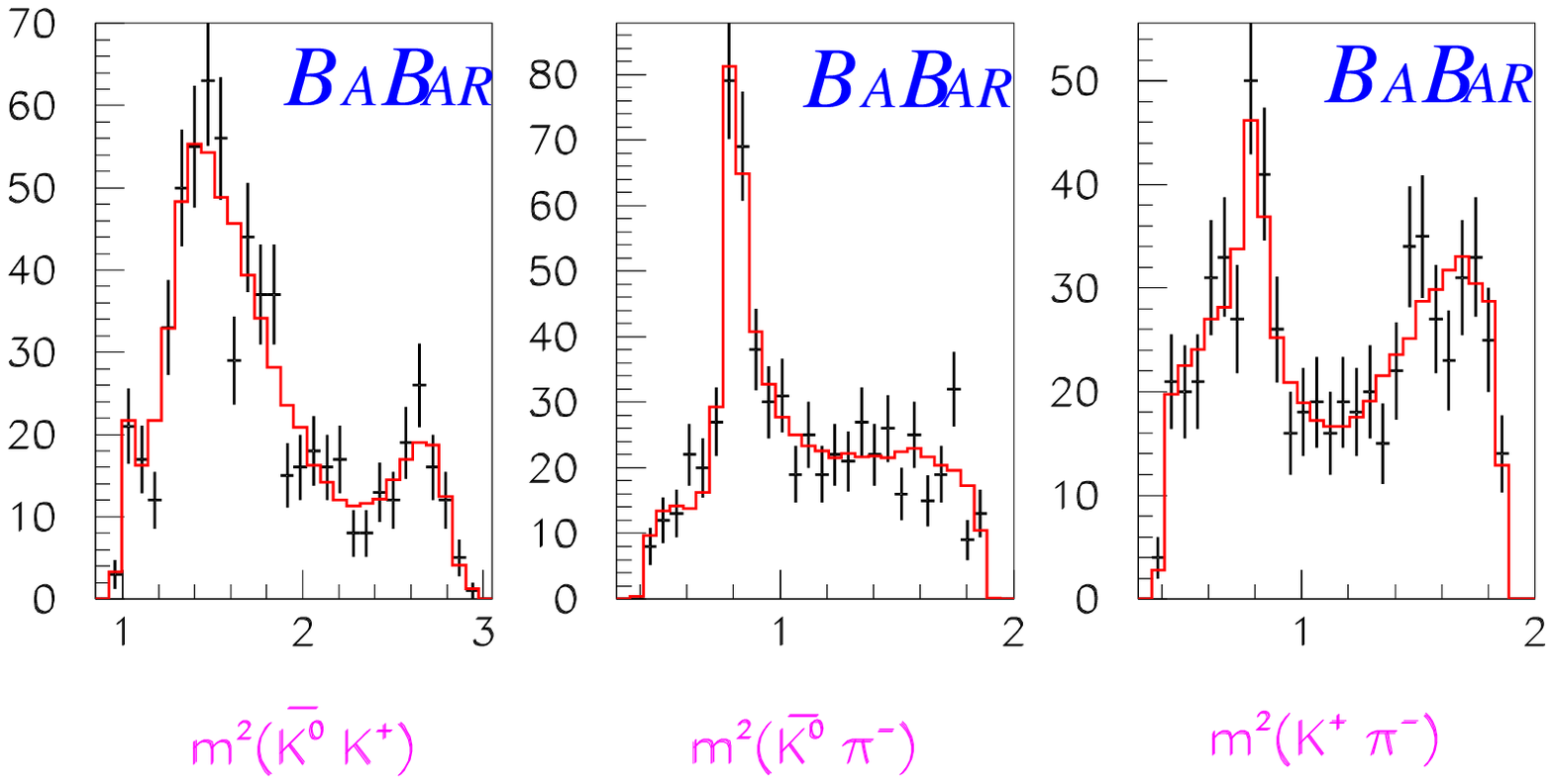}
\caption{Dalitz plot projections for  $D^0 \to \overline{K}^0 K^+ \pi^-$. 
The data are
represented with error bars; the histogram is 
a projection of the fit described in the text.}
\label{fig:pro_k0skpi2}
\end{center}
\end{figure}

The Dalitz plot for $D^0 \to \overline{K}^0 K^+ \pi^-$ candidates is shown in Fig.~\ref{fig:dalitz_k0skpi_2} 
and its projections are shown in Fig.~\ref{fig:pro_k0skpi2}.
The presence of the $K^{*+}(892)$ resonance is evident together with a small 
$K^{*0}(892)$ contribution with asymmetric lobes suggesting interference
with other final states. A shoulder near the threshold of the $\overline{K}^0 K^+$ 
mass distribution suggests the presence of the $a_0^+(980)$ resonance.
This $D^0$ decay channel is fit using the final states  
shown in 
Table~\ref{tab:res_k0skpib}, which also displays the fit fractions and 
phases. The resulting $\chi^2$ for this fit is given in 
Table~\ref{tab:all}.

\begin{table}[htbp]
\caption{
Preliminary results from the Dalitz plot analysis of $D^0 \to \overline{K}^0 K^+ \pi^-$.
}
\label{tab:res_k0skpib}
\begin{center}
\vskip -0.2cm
\begin{tabular}{lcc}
\hline
 Final state & Fraction (\%) & Phase (degrees)\cr
\hline
& & \\
 $ K^{*0}_0(1430) \overline{K}^0$ & 26.0 $\pm$ 16.1 $\pm$ 3.3 &--38 $\pm$ 22\\
 $ K^{*0}_1(892) \overline{K}^0$ &  2.8 $\pm$ 1.4 $\pm$ 0.5 &--126 $\pm$ 19\\
 $ K^{*0}_1(1680) \overline{K}^0$ &15.2 $\pm$ 11.9 $\pm$  0.5& 161 $\pm$ 9 \\
 $ K^{*0}_2(1430) \overline{K}^0$ &1.7 $\pm$ 2.5 $\pm$ 0.2 & 53 $\pm$ 38\\
 $K^{*-}_0(1430) K^+$ & 2.4 $\pm$ 8.2 $\pm$ 1.0 &--142 $\pm$ 115\\
 $K^{*-}_1(892) K^+$ & 35.6 $\pm$ 7.7 $\pm$ 2.3 & 0 \\ 
 $K^{*-}_1(1680) K^+$ & 5.1 $\pm$ 5.7 $\pm$ 1.1 &124 $\pm$ 27 \\
 $K^{*-}_2(1430) K^+$ & 1.0 $\pm$ 1.0 $\pm$ 0.2 &--26 $\pm$ 38 \\
 $a_0^+(980) \pi^-$ & 15.1 $\pm$ 12.5 $\pm$ 0.6 & -160 $\pm$ 42\\
 $a_0^+(1450) \pi^-$ & 2.2 $\pm$ 2.7 $\pm$ 1.2 & 148 $\pm$ 25\\
 N.R. & 36.6 $\pm$ 25.8 $\pm$ 2.7& -172 $\pm$ 13 \\
& & \\
\hline
Sum &  144 $\pm$ 37 & \\
\hline
\end{tabular}
\end{center}
\end{table}

This fit reveals a rather complex structure with large uncertainties
due to 
the limited sample size. A sizeable nonresonant contribution is predicted by
the fit,
in contrast to the $D^0 \to K^0 K^- \pi^+$ channel discussed
in the last section.

Dropping the nonresonant contribution results in a decrease of $2 \log L$ of 9 units
for 2 less parameters with the same $\chi^2$ value. While all other fractions have little
changes, in this fit the contribution from  $K^{*-}_0(1430) K^+$ 
increases 
from $(2\pm8)$\% to $(26 \pm 12)$\% and that of $a_0^+(980) \pi^-$ 
decreases from $(15\pm13)$\% to 
$(5 \pm 6)$\%.
The sum of the fractions for this fit is (114 $\pm$ 23)\%.

\section{\boldmath $D^0 \to \overline{K}^0 K^+ K^-$ Dalitz Plot Analysis}

\begin{figure}[!htb]
\begin{center}
\includegraphics[height=12cm,width=12cm]{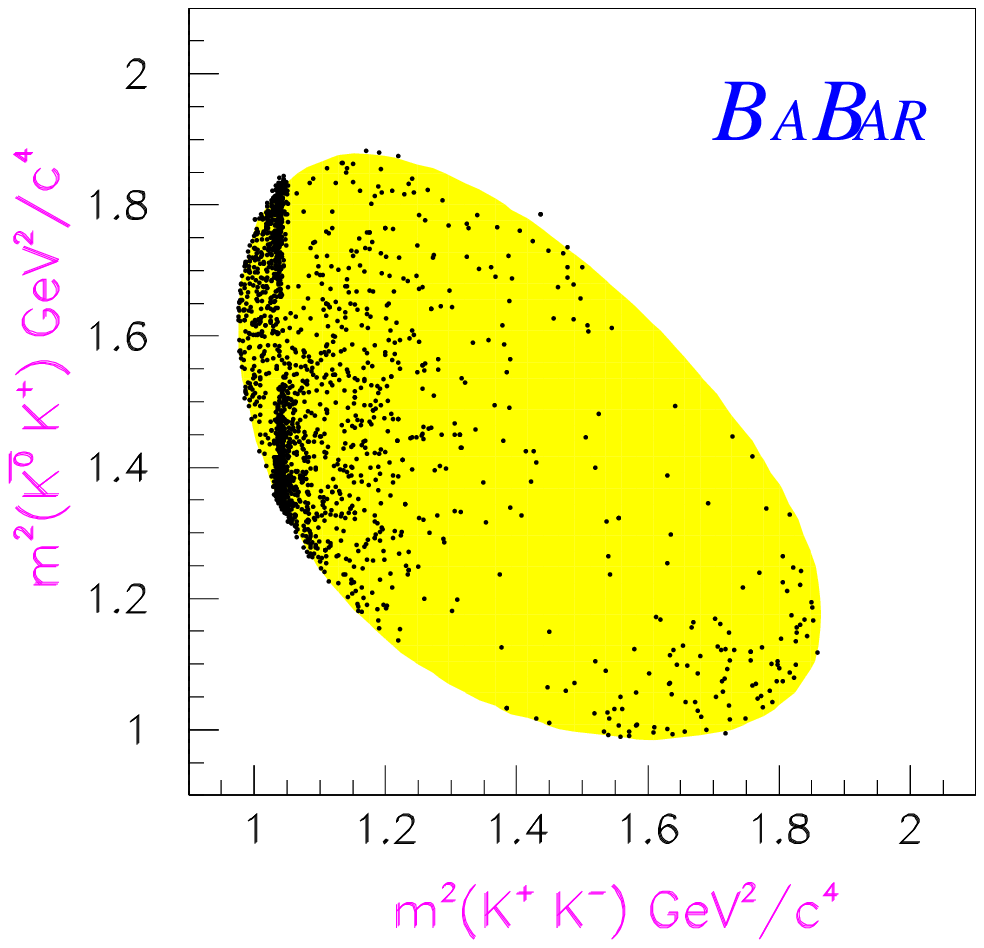}
\caption{Dalitz plot of $D^0 \to \overline{K}^0 K^+ K^-$.
} 
\label{fig:dalitz_k0skk}
\end{center}
\end{figure}

\begin{figure}[!htb]
\begin{center}
\includegraphics[height=8cm,width=16cm]{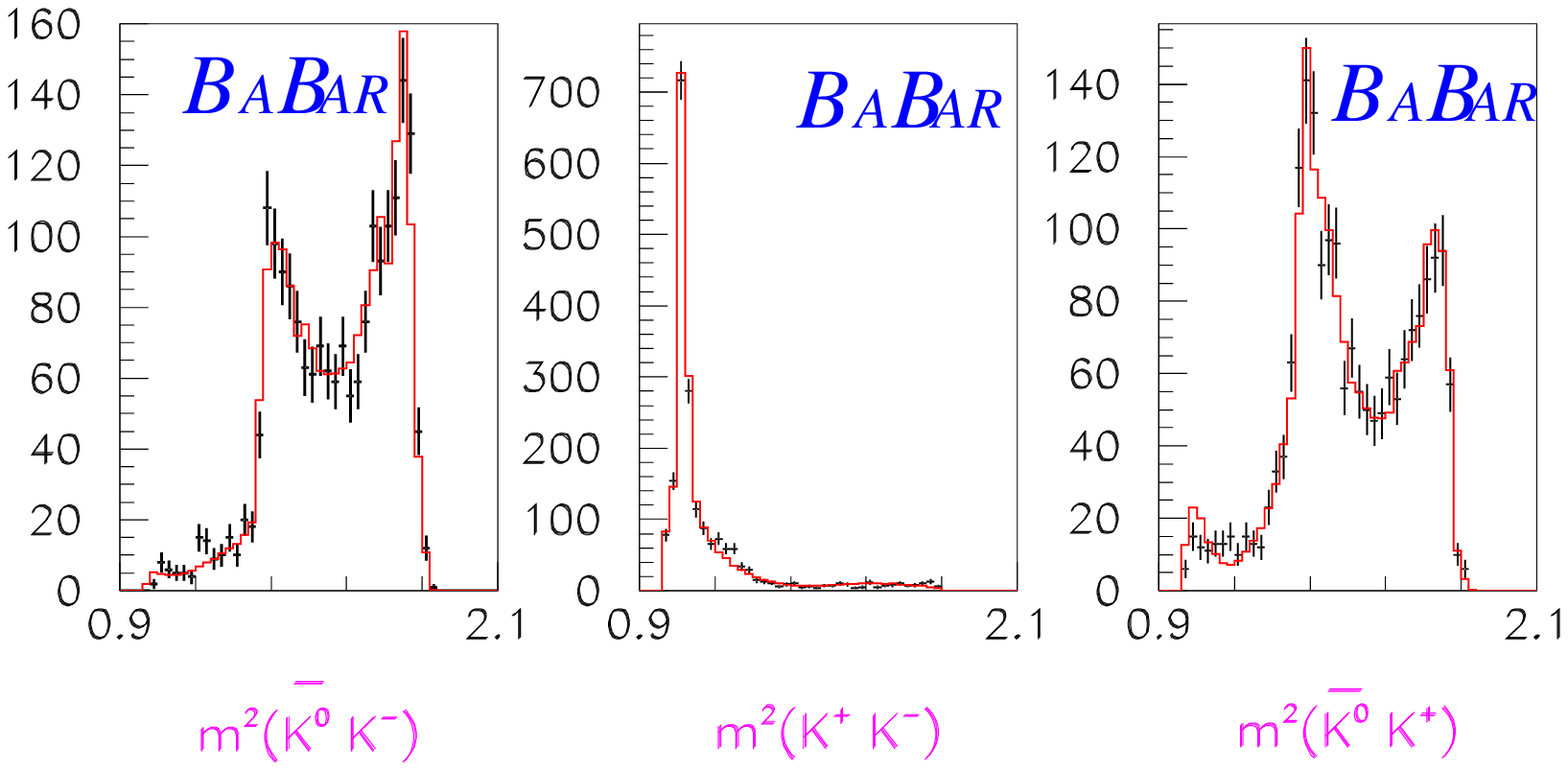}
\caption{Dalitz plot projections for  $D^0 \to \overline{K}^0 K^+ K^-$. 
The data are
represented with error bars; the histogram is 
the projection of the fit described in the text.}
\label{fig:pro_k0skk}
\end{center}
\end{figure}

The Dalitz plot for $D^0 \to \overline{K}^0 K^+ K^-$ candidates is shown 
in Fig.~\ref{fig:dalitz_k0skk}
and its projections are shown in Fig.~\ref{fig:pro_k0skk}.
A strong $\phi(1020)$ signal interfering with a threshold
scalar meson can be clearly seen. 
Both $f_0(980)$ and $a_0^0(980)$ resonances can be present near threshold.
An accumulation of 
events due to a charged $a_0^+(980)$ can be observed on the lower edge 
of the Dalitz plot.
Table~\ref{tab:res_k0skk} shows the list of the final states and the fitted
fractions and phases for this $D^0$ decay channel. 

\begin{table}[htbp]
\caption{
Preliminary results from the Dalitz plot analysis of $D^0 \to \overline{K}^0 K^+ K^-$.
}
\label{tab:res_k0skk}
\begin{center}
\vskip -0.2cm
\begin{tabular}{lcc}
\hline
Final state &  Fraction (\%) & Phase (degrees)\cr
\hline
& & \\
$\overline{K}^0 \phi$& 45.4 $\pm$ 1.6 $\pm$ 1.0 & 0 \\
$\overline{K}^0 a_0^0(980)$& 60.9 $\pm$ 7.5 $\pm$ 13.3 & 109 $\pm$ 5\\
$\overline{K}^0 f_0(980)$ & 12.2 $\pm$  3.1 $\pm$ 8.6 & --161 $\pm$ 14  \\
$a_0(980)^+ K^-$& 34.3 $\pm$ 3.2 $\pm$ 6.8 & --53 $\pm$ 4 \\
$a_0(980)^- K^+$& 3.2 $\pm$ 1.9 $\pm$ 0.5 & -13 $\pm$ 15 \\
N.R.& 0.4 $\pm$ 0.3 $\pm$ 0.8 & 40 $\pm$ 44 \\
& & \\
\hline
Sum &  156 $\pm$ 9 & \\
\hline
\end{tabular}
\end{center}
\end{table}

Most of the uncertainty in this channel is due to the poorly defined parameters
of $a_0(980)$ and $f_0(980)$. Their properties are taken from 
measurements performed by other experiments. 
Since these states lie below $K \Kbar$ threshold, their properties
cannot be measured in this $D^0$ decay mode only. 
Their line shape would be best
determined from a coupled-channel analysis of $K^0_S K^0_S$, $K^0_S K^0_L$, 
$K^+ K^-$, $\eta \pi$ and $\pi \pi$ decays.
A scan is made of the likelihood function with respect to the ratio
 $g_{K \Kbar}/g_{\eta \pi}$ for $a_0(980)$ leaving the other parameters fixed.
The likelihood has a maximum for $g_{K \Kbar}/g_{\eta \pi}\sim$1.3, to be compared
with the Crystal Barrel result of $1.01\pm 0.07$.

Similarly, a scan is performed as a function of $g_K/g_{\pi}$ for $f_0(980)$.
The likelihood has a maximum for $g_K/g_{\pi}\sim$1.25, to be compared with
the WA76 result of $2.0\pm 0.6$. The variation of the fractions
due to the change of these parameters is taken into account in the evaluation of the 
systematic errors.

The doubly-Cabibbo-suppressed contribution $D^0 \to K^0 K^+ K^-$
has also been included in the fit. The presence of such a contribution should show up as an $a_0^-(980)$ signal in the
$K^0 K^-$ effective mass. The insertion of this new term results in an increase
of $2 \log L$  by 10 units for 2 additional parameters
 and the resulting $\chi^2$ is included in Table~\ref{tab:all}. 
However, with the
present statistics, its contribution of (3.2 $\pm$ 1.9)\% is consistent 
with zero.

The poor fit quality for this channel, indicated by the
$\chi^2/N_{cells}$ of $98/59$
in Table~\ref{tab:all}, comes mostly 
from the $\phi$ mass region.  Effects of $\phi$ mass resolution comparable 
to its natural width combined with uncertainties in the $a_0$ and $f_0$ line
shapes are not included in the current fit model, and do not
correctly model the interference in this region.  Future
analyses with larger data samples should attempt to address this
problem with a more detailed model.

\section{Summary}
\label{sec:Summary}

Dalitz plot analyses are performed for the hadronic 
decays $D^0 \to K^0 K^- \pi^+$, $D^0 \to \overline{K}^0 K^+ \pi^-$ and 
$D^0 \to \overline{K}^0 K^+ K^-$.
The fractions and relative phases for intermediate resonant states are extracted.
The following preliminary ratios of branching fractions are measured: 
$$\frac{D^0 \to K^0 K^- \pi^+}{D^0 \to \overline{K}^0 \pi^+ \pi^-} = (8.32 \pm 0.29 (stat) \pm 0.56(syst)) \times 10^{-2},$$
$$\frac{D^0 \to \overline{K}^0 K^+ \pi^-}{D^0 \to \overline{K}^0 \pi^+ \pi^-} = (5.68 \pm 0.25(stat) \pm 0.41(syst)) \times 10^{-2},$$
$$\frac{D^0 \to \overline{K}^0 K^+ K^-}{D^0 \to \overline{K}^0 \pi^+ \pi^-} = (16.30 \pm 0.37(stat) \pm 0.27(syst)) \times 10^{-2}.$$

\section{Acknowledgments}
\label{sec:Acknowledgments}

% Specific acknowledgments for this paper; remove if not needed.

% Standard acknowledgments paragraph; must always be included.
\input pubboard/acknowledgements

\end{document}

%% file: pubboard/authors_ICHEP2002.tex
\begin{center}
\small

The \babar\ Collaboration,
\bigskip

%% author list as of 05-Jul-2002 (556 authors)
B.~Aubert,
D.~Boutigny,
J.-M.~Gaillard,
A.~Hicheur,
Y.~Karyotakis,
J.~P.~Lees,
P.~Robbe,
V.~Tisserand,
A.~Zghiche
\inst{Laboratoire de Physique des Particules, F-74941 Annecy-le-Vieux, France }
A.~Palano,
A.~Pompili
\inst{Universit\`a di Bari, Dipartimento di Fisica and INFN, I-70126 Bari, Italy }
% ...

\end{center}\newpage

%% file: pubboard/acknowledgements.tex
We are grateful for the 
extraordinary contributions of our \pep2\ colleagues in
achieving the excellent luminosity and machine conditions
that have made this work possible.
The success of this project also relies critically on the 
expertise and dedication of the computing organizations that 
support \babar.
The collaborating institutions wish to thank 
SLAC for its support and the kind hospitality extended to them. 
This work is supported by the
US Department of Energy
and National Science Foundation, the
Natural Sciences and Engineering Research Council (Canada),
Institute of High Energy Physics (China), the
Commissariat \`a l'Energie Atomique and
Institut National de Physique Nucl\'eaire et de Physique des Particules
(France), the
Bundesministerium f\"ur Bildung und Forschung and
Deutsche Forschungsgemeinschaft
(Germany), the
Istituto Nazionale di Fisica Nucleare (Italy),
the Research Council of Norway, the
Ministry of Science and Technology of the Russian Federation, and the
Particle Physics and Astronomy Research Council (United Kingdom). 
Individuals have received support from 
the A. P. Sloan Foundation, 
the Research Corporation,
and the Alexander von Humboldt Foundation.